\documentstyle[12pt]{article}

\def\Journal#1#2#3#4{{#1} {\bf #2}, #3 (#4)}

\def\PLB{{\em Phys. Lett.}  B}

\def\PRD{{\em Phys. Rev.} D}

\def\be{\begin{equation}}
\def\ee{\end{equation}}
\def\bea{\begin{eqnarray}}
\def\eea{\end{eqnarray}}

\begin{document}
\baselineskip=16pt
\begin{titlepage}
\rightline{hep-ph/9909535}
\rightline{September 1999}  
\begin{center}

\vspace{1cm}

\large {\bf 5-Dimensional Assisted Inflation and the Remedy of the
Fine-Tuning Problem}~\footnote{Talk presented at the ``Workshop on
Current Issues in String Cosmology", June 21-25, 1999, IHES, Paris
and at the ``XIth Rencontres de Blois: Frontiers
of Matter", June 27-July 3, 1999, Blois, France.}
\vspace*{10mm}
\normalsize

{\bf Panagiota Kanti}  
 
\smallskip 
\medskip 
 
{\it Theoretical Physics Institute, School of Physics and
Astronomy,\\[1mm] University of Minnesota, Minneapolis, MN 55455, USA} 
\end{center} 
\vskip0.8in 
\centerline{\large\bf Abstract}
\vspace*{4mm} 
We extend the idea of assisted inflation to the case of power-law
potentials and demonstrate the simultaneous resolution of two
major problems that plague chaotic inflation. The implementation of the
same idea in the framework of a 5-dimensional, scalar field theory
leads to a model of chaotic inflation free of fine-tuning.

\vspace*{20mm}
\hfill
\end{titlepage}

\section{Introduction}

The chaotic inflationary scenario~\cite{linde} is undoubtly the simplest of
all the available models of inflation. However, it suffers from two major
shortcomings: the large initial conditions on the inflaton field 
($\tilde{\phi}_0 > {\rm few} M_P$) and the severe fine-tuning of the
coupling parameter ($\lambda \sim 10^{-12}$). Here, we will 
resolve both of the above problems by implementing the idea of assisted
inflation, first, in the framework of a 4-dimensional
theory of multiple scalar fields, and secondly, in the context of a
5-dimensional theory of a single, self-interacting, scalar field~\cite{ko1}.

\section{Assisted Inflation in 4 Dimensions}

We start by presenting, in its simplest form, the idea of assisted
inflation as originally proposed by Liddle, Mazumdar and Schunck~\cite{liddle}
and studied further by Malik and Wands~\cite{wands}. We consider a
massless, scalar field with a self-interacting potential of the form
\be
V=V_0\,e^{-\sqrt{2 \over p}\,\phi}\,,
\ee
where $V_0$ is a constant, and $p$ the slope parameter of the potential.
The above scalar theory leads to a power-law expansion of the universe,
i.e. $R(t) \sim t^p$, and if $p>1$, the expansion is fast enough
to resolve the problems of the Standard Cosmological Model. On
the other hand, we may assume the existence of multiple, scalar fields
$\phi_i$, $i=1,...,N$, with the same type of potential and, by recognizing
the late-time attractor of the system, i.e. the configuration of the fields
that minimizes the  potential energy, we can map the original theory of
multiple fields to a theory of a single field with the same type of
self-interaction. This effective theory leads once again to a power-law
expansion of the universe with a power $\tilde{p}=N\,p$, which for large
$N$ gives an adequate amount of inflation even for $1/3<p<1$.

Here, we will be mainly interested in power-law potentials which are
easily applicable to chaotic inflation. Therefore, we consider the
following theory of multiple fields
\be
-{\cal L} = \sum_{i=1}^N \Biggl\{\frac{(\partial \phi_i)^2}{2}
+  \frac{m^2}{2}\,\phi_i^2 + 
\frac{\lambda}{4!}\,\phi_i^4 \Biggl\}\,.
\ee
By making use of the late-time attractor, which has all the fields equal,
the above theory can be mapped to a theory of a single field
-- the inflaton -- by making the redefinitions 
\be
\tilde{\phi}=\sqrt{N}\,\phi_i\,, \qquad
\tilde{\lambda}= \frac{\lambda}{N}\,.
\ee
Note, that the above theory can serve as a model for chaotic inflation
which, quite remarkably, is free of its two major problems. For a large
number of fields $N$, the inflaton
field $\tilde{\phi}$ can easily reach the value of a few $M_P$ while the
values of $\phi_i$ can be well below the above threshold. In addition,
the quartic coupling parameter naturally acquires an extremely small value,
thus, resolving the problem of fine tuning. However, for the assistance
method to work, the assisted sector needs to be non-coupled, otherwise,
the presence of multiple fields impedes inflation since it leads to
$\tilde{p}=p/N$, in the former case~\footnote{A remedy, in the case of
coupled scalar fields with exponential potentials, has been proposed by
Copeland, Mazumdar and Nunes~\cite{copeland}.}, and
$\tilde{\lambda}=\lambda\,N^2$, in the latter case.

\section{Assisted Inflation in 5 Dimensions}

As a possible source of a theory with a large number of scalar fields,
we consider the following 5-dimensional theory of a self-interacting,
massless scalar field
\be
S_5= -\int d^4x\,dz\,\sqrt{G_5}\,\Biggl\{\frac{M_5^3}{16\pi}\,R_5
+\frac{1}{2}\,G^{AB}_5\,\partial_A \hat{\phi}\, \partial_B \hat{\phi}
+\frac{\hat{\lambda}}{4!}\,\frac{\hat{\phi}^4}{M_5}\Biggr\}\,,
\ee
where $\hat{\lambda}$ is the 5-dimensional quartic coupling constant
and $M_5$ the 5-dimensional Planck mass. We assume that the extra
dimension is compactified along a circle with circumference $2L$ and
we Fourier expand $\hat{\phi}$ as
\be
\hat \phi(x,z) = \hat \phi_0(x) + \sum_{n = 1}^{N} 
\,\Bigl[ \hat \phi_n(x)\,e^{i\frac{n\pi}{L}z} + 
\hat \phi^{*}_n(x)\,e^{-i \frac{n\pi}{L}z} \Bigr]\,,
\ee
where $N \sim 2L M_5=(M_P/M_5)^2$ is the maximum number of Kaluza-Klein
(KK) states. Then, the 5-dimensional, fundamental theory is reduced to a
4-dimensional, effective one which has the form
\bea
&~& \hspace*{-0.5cm} 
S_4 = -\int d^4x\,\sqrt{g}\,\Biggl\{\frac{M_P^2}{16\pi}\,R+
\frac{(\partial \phi_0)^2}{2}+ 
\sum^{N}_{n=1} \Bigl(|\partial\phi_n|^2 +
\frac{n^2\pi^2}{L^2}\,|\phi_n|^2 \Bigr) +
\frac{\lambda}{4!}\,\biggl[\phi^4_0 + 12 \phi^2_0 
\sum_{n=1}^{N} \phi_n \phi_n^*\nonumber \\[3mm]
&~& \hspace*{0.5cm} +\,12 \phi_0 \sum_{n,k=1}^{N} 
\Bigl(\phi_n \phi_k \phi^*_{n+k}+ h.c.\Bigr)+\sum_{n,k,l =1}^{N} 
\Bigl( 4 \phi_n \phi_k \phi_l \phi^*_{n+k+l} + h.c.  +
6 \phi_n \phi_k \phi^*_l \phi^*_{n+k-l} \Bigr)\biggr]\Biggl\}\,\,,
\nonumber
\eea
and where the following redefinitions have been used~\footnote{We assume that
$G_{55}=e^{2\gamma}$ is fixed and we ignore the KK gauge field
$G_{\mu 5}=e^{2\gamma}A_\mu$.}
\be
M_P^2=2L\, M_5^3\,,\qquad \phi_i=\sqrt{2L}\,\hat{\phi}_i\,,
\qquad \lambda=\frac{\hat{\lambda}}{2L M_5}\simeq 
\frac{\hat{\lambda}}{N}
\ee
Note that, for large $N$, the 4-dimensional fields $\phi_i$ are considerably
less strongly coupled than the original 5-dimensional one.
 
In order to map the above system of multiple, heavily-coupled scalar
fields to a theory of a single, self-interacting field, which is known to
lead to chaotic inflation, we need to determine the late-time attractor
of the system. For that purpose, we make the following
simplifications and assumptions: 
\begin{enumerate}
\item We substitute the $N$ complex KK fields with $2N$ real KK fields.
\item We impose the periodic condition $\phi_{2N+n}=\phi_n$ in order to
resolve the asymmetry with which the two ``boundary fields" $\phi_1$ and
$\phi_{2N}$ enter the Lagrangian.
\item  We assume that $m_n^2 << \lambda \phi_0^2/2$,
i.e. that the KK masses can been considered negligible.
\end{enumerate}
Then, the late-time
attractor is easily determined and is found to be the unique one and to
have all the fields equal, thus, reducing the theory of multiple fields
to the following one:
\be
-{\cal L}_{4D} = \frac 12\,(\partial \tilde{\phi})^2 +
A(N,q)\, \frac{\lambda}{4!}\,\tilde{\phi}^4\,,
\ee
where, now, the inflaton field is defined as
\be 
\tilde{\phi}=\sqrt{1+\frac{2N}{q^2}}\,\phi_0\,, \qquad \phi_0=q\,\phi_n\,.
\ee
Note that, apart from the coefficient $A(N,q)$, the inflaton field
is a self-interacting field with a quartic potential. The system of
equations of motion accepts two solutions for the proportionality
coefficient which give $q \sim N$ and lead to $A(N,q) \sim 1$
and, thus, to a model of chaotic inflation where the inflaton field is
significantly more weakly coupled than the original, 5-dimensional
one~\footnote{There is another solution for $q$, namely $q= const$, which
leads to $A(N,q) \sim N$ and, thus, to an inflaton field which is equally
strongly coupled compared to the 5-dimensional field $\hat{\phi}$.}.
This is exactly the type of solution we were searching for. If we
assume that $\hat V(\hat \phi) \sim
M_5^5$, we obtain the 4-dimensional constraint $\tilde V(\tilde \phi)
\sim \lambda\,\tilde{\phi}^4 \sim M_P^2 M_5^2$. Substituting the values
$\lambda \sim 10^{-12}$ and $\tilde{\phi} \sim M_P$, which are necessary 
for the occurrence of chaotic inflation, we are led to the constraint
$M_5 \geq 10^{-6}\,M_P$ and subsequently to $\lambda \geq
10^{-12}\,\hat{\lambda}$. 

We may, then, conclude that by starting with
a 5-dimensional, scalar field theory with a coupling parameter
of ${\cal O}(1)$, we obtain a 4-dimensional, effective theory
of a single, self-interacting scalar field which may have a coupling
parameter as small as $10^{-12}$ without the need of any fine-tuning.
Unfortunately, the above model does not lead to the relaxation of the
large initial conditions on the inflaton field, however, a similar
theory based on a 5-dimensional, non-interacting, massive
field~\cite{ko2} can provide a remedy for this problem, too.

\section*{Acknowledgments}
This work was done in collaboration with K.A. Olive and was
supported in part by U.S. DOE Grant No. DE-FG02-94ER40823
at Minnesota.

\end{document}